\pdfoutput=1
\documentclass[epjc3]{svjour3}
\usepackage{ifthen} % for conditional statements
\newboolean{pdflatex}
\setboolean{pdflatex}{true} % False for eps figures
 
\newboolean{articletitles}
\setboolean{articletitles}{true} % False removes titles in references
 
\newboolean{uprightparticles}
\setboolean{uprightparticles}{false} %True for upright particle symbols
 
\newboolean{inbibliography}
\setboolean{inbibliography}{false} %True once you enter the bibliography
 
\newcommand{\bdecay}{\ensuremath{\mbox{B}^+ \rightarrow p \pi^+ \pi^+
\overline{\Sigma}_c^{--}}\xspace}
\newcommand{\bdecaychain}{\ensuremath{\mbox{B}^+ \rightarrow p \pi^+ \pi^+
\big(\overline{\Sigma}_c^{--} \xspace \rightarrow \overline{\Lambda}_c^- \pi^-\big)} \xspace}
\newcommand{\ldecay}{\ensuremath{{\mathrm \Lambda}_c \rightarrow p K \pi} \xspace}
\newcommand{\nonresdecay}{\ensuremath{\mbox{B}^+ \rightarrow p \pi^+ \pi^+
\pi^-\overline{\Lambda}_c^{-}}\xspace}

\newcommand{\bodecay}{\ensuremath{\mbox{B}^\circ \rightarrow p \pi^+ \pi^+
\pi^-\pi^-\overline{\Lambda}_c^{-}}\xspace}

\newcommand{\nonresdecpi}{\ensuremath{\mbox{B}^+ \rightarrow p \pi^+ \pi^+
\pi^-\overline{\Lambda}_c^{-} \pi^0} \xspace}
 
\usepackage{graphicx}  % to include figures (can also use other packages)
\usepackage {array}
\usepackage{microtype}
\usepackage{lineno}  % for line numbering during review
\usepackage{xspace} % To avoid problems with missing or double spaces after
\usepackage{graphicx}  % to include figures (can also use other packages)
\usepackage{color}
\usepackage{colortbl}
\usepackage{amsmath} % Adds a large collection of math symbols
\usepackage{amssymb}
\usepackage{amsfonts}
\usepackage{upgreek} % Adds in support for greek letters in roman typeset

\usepackage{booktabs}
\usepackage{multirow}
\usepackage{subfigure}
\newcommand*\patchAmsMathEnvironmentForLineno[1]{%
\expandafter\let\csname old#1\expandafter\endcsname\csname #1\endcsname
\expandafter\let\csname oldend#1\expandafter\endcsname\csname
end#1\endcsname
 \renewenvironment{#1}%
   {\linenomath\csname old#1\endcsname}%
   {\csname oldend#1\endcsname\endlinenomath}%
}
\newcommand*\patchBothAmsMathEnvironmentsForLineno[1]{%
  \patchAmsMathEnvironmentForLineno{#1}%
  \patchAmsMathEnvironmentForLineno{#1*}%
}
\AtBeginDocument{%
\patchBothAmsMathEnvironmentsForLineno{equation}%
\patchBothAmsMathEnvironmentsForLineno{align}%
\patchBothAmsMathEnvironmentsForLineno{flalign}%
\patchBothAmsMathEnvironmentsForLineno{alignat}%
\patchBothAmsMathEnvironmentsForLineno{gather}%
\patchBothAmsMathEnvironmentsForLineno{multline}%
}
 
\usepackage{hyperref}    % Hyperlinks in references
\usepackage[all]{hypcap} % Internal hyperlinks to floats.
 
\def\ux85 {UX85\xspace}

\ifthenelse{\boolean{uprightparticles}}%
{

 \def\PDelta      {\ensuremath{\Delta}\xspace}                 
 \def\PXi      {\ensuremath{\Xi}\xspace}                 
 \def\PLambda      {\ensuremath{\Lambda}\xspace}                 
 \def\PSigma      {\ensuremath{\Sigma}\xspace}                 
 \def\POmega      {\ensuremath{\Omega}\xspace}                 
 \def\PUpsilon      {\ensuremath{\Upsilon}\xspace}                 
 
 \def\PB      {\ensuremath{\mathrm{B}}\xspace}                 
                  
 \def\PD      {\ensuremath{\mathrm{D}}\xspace}

 \def\PK      {\ensuremath{\mathrm{K}}\xspace}

 \def\Pi      {\ensuremath{\mathrm{i}}\xspace}

}
{

 \mathchardef\PDelta="7101
 \mathchardef\PXi="7104
 \mathchardef\PLambda="7103
 \mathchardef\PSigma="7106
 \mathchardef\POmega="710A
 \mathchardef\PUpsilon="7107
                  
 \def\PB      {\ensuremath{B}\xspace}                 
                  
 \def\PD      {\ensuremath{D}\xspace}

 \def\PK      {\ensuremath{K}\xspace}

 \def\Pi      {\ensuremath{i}\xspace}

}

\def\kaon  {\ensuremath{\PK}\xspace}
  \def\Kbar  {\kern 0.2em\overline{\kern -0.2em \PK}{}\xspace}

\def\Kz    {\ensuremath{\kaon^0}\xspace}
\def\Kzb   {\ensuremath{\Kbar^0}\xspace}
\def\KzKzb {\ensuremath{\Kz \kern -0.16em \Kzb}\xspace}
\def\Kp    {\ensuremath{\kaon^+}\xspace}
\def\Km    {\ensuremath{\kaon^-}\xspace}

\def\KpKm  {\ensuremath{\Kp \kern -0.16em \Km}\xspace}

  \def\Dbar    {\kern 0.2em\overline{\kern -0.2em \PD}{}\xspace}
\def\D       {\ensuremath{\PD}\xspace}

\def\Dz      {\ensuremath{\D^0}\xspace}
\def\Dzb     {\ensuremath{\Dbar^0}\xspace}
\def\DzDzb   {\ensuremath{\Dz {\kern -0.16em \Dzb}}\xspace}
\def\Dp      {\ensuremath{\D^+}\xspace}
\def\Dm      {\ensuremath{\D^-}\xspace}

\def\DpDm    {\ensuremath{\Dp {\kern -0.16em \Dm}}\xspace}

  \def\Bbar    {\kern 0.18em\overline{\kern -0.18em \PB}{}\xspace}

  \def\Y#1S{\ensuremath{\PUpsilon{(#1S)}}\xspace}% no space before {...}!

\def\AT#1     {\ensuremath{A_{\mathrm{T}}^{#1}}\xspace}           % 2

\def\C#1      {\ensuremath{\mathcal{C}_{#1}}\xspace}                       % 9
\def\Cp#1     {\ensuremath{\mathcal{C}_{#1}^{'}}\xspace}                    % 7
\def\Ceff#1   {\ensuremath{\mathcal{C}_{#1}^{\mathrm{(eff)}}}\xspace}        % 9  
\def\Cpeff#1  {\ensuremath{\mathcal{C}_{#1}^{'\mathrm{(eff)}}}\xspace}       % 7
\def\Ope#1    {\ensuremath{\mathcal{O}_{#1}}\xspace}                       % 2
\def\Opep#1   {\ensuremath{\mathcal{O}_{#1}^{'}}\xspace}                    % 7

\newcommand{\tev}{\ensuremath{\mathrm{\,Te\kern -0.1em V}}\xspace}
\newcommand{\gev}{\ensuremath{\mathrm{\,Ge\kern -0.1em V}}\xspace}
\newcommand{\mev}{\ensuremath{\mathrm{\,Me\kern -0.1em V}}\xspace}
\newcommand{\kev}{\ensuremath{\mathrm{\,ke\kern -0.1em V}}\xspace}
\newcommand{\ev}{\ensuremath{\mathrm{\,e\kern -0.1em V}}\xspace}
\newcommand{\gevc}{\ensuremath{{\mathrm{\,Ge\kern -0.1em V\!/}c}}\xspace}
\newcommand{\mevc}{\ensuremath{{\mathrm{\,Me\kern -0.1em V\!/}c}}\xspace}
\newcommand{\gevcc}{\ensuremath{{\mathrm{\,Ge\kern -0.1em V\!/}c^2}}\xspace}
\newcommand{\gevgevcccc}{\ensuremath{{\mathrm{\,Ge\kern -0.1em V^2\!/}c^4}}\xspace}
\newcommand{\mevcc}{\ensuremath{{\mathrm{\,Me\kern -0.1em V\!/}c^2}}\xspace}

\def\gsim{{~\raise.15em\hbox{$>$}\kern-.85em
          \lower.35em\hbox{$\sim$}~}\xspace}
\def\lsim{{~\raise.15em\hbox{$<$}\kern-.85em
          \lower.35em\hbox{$\sim$}~}\xspace}

\def\tell1  {TELL1\xspace}
\def\ukl1   {UKL1\xspace}

\usepackage{cite} % Allows for ranges in citations
\begin{document}

% Title --------------------------------------------------
\title{A method to measure the absolute branching fractions of $\Lambda_c$ decays}

\vspace*{2.0cm}

% Authors -------------------------------------------------
\author{A.~Contu, D.~Fonnesu, R.G.C.~Oldeman, B.~Saitta and C.~Vacca}

\institute {Dipartimento di Fisica Universit\`a di Cagliari and INFN, Sezione di Cagliari, Italy}

\date {Received: 28 August 2014}

\maketitle

% Abstract -----------------------------------------------
\begin{abstract}
 \noindent
It is proposed to exploit the decay 
of the meson 
\bdecay 
and of its charge conjugate $B^-$
copiously produced at LHC to obtain
a sample of $\Lambda_c$ baryons through the strong decay 
$\Sigma_c \rightarrow \Lambda_c \pi$.
The sample thus obtained is not affected by biases typically
introduced by selections that depend on
specific 
decay modes. Therefore it allows a measurement 
of the absolute
branching fraction for the decay of the $\Lambda_c$ baryon into $ p K \pi$
or into
other observable final states to be performed in a model independent manner. 
The accuracy that
can be achieved with this method is discussed and it is 
shown that it would be either competitive with
or an improvement over current measurements.
\end{abstract}

\pagestyle{empty}  % no page number for the title 

%%%%%%%%%%%%%%%%%%%%%%%%%%%%%%%%
%%%%%  EOD OF TITLE PAGE  %%%%%%
%%%%%%%%%%%%%%%%%%%%%%%%%%%%%%%%

%  empty page follows the title page ----

\setcounter{page}{2}
\mbox{~}

%%%%%%%%%%%%%%%%%%%%%%%%%
%%%%% Main text %%%%%%%%%
%%%%%%%%%%%%%%%%%%%%%%%%%

\pagestyle{plain} % restore page numbers for the main text
\setcounter{page}{1}
\pagenumbering{arabic}

% %%%%%%% CHOOSE --------
%% ----------------------------------
%% Line numbering on the left margin 
%% ----------------------------------
%% Uncomment during review phase. 
%% Comment it out before a final submission.
%\linenumbers
%% --------------------------------
% %%%%%%%%%%%%% ---------

\section{Introduction}

 Recently the Belle collaboration \cite{Belle} has reported a value
of $(6.84\pm0.24 ^{+0.21}_{-0.27})\%$ for
the absolute branching 
fraction of the decay \ldecay,
obtained from the reconstruction of the system
$D^* \bar{p} \pi$ recoiling against the $\Lambda_c$ produced in 
$e^+e^-$ annihilation. This measurement is model independent and
has a significantly better precision than earlier results 
by the CLEO \cite{Cleo} and the ARGUS \cite{Argus} 
collaborations, which
were deduced making model dependent assumptions and
are marginally consistent with one another. 

In this paper it is suggested to exploit a particular 
decay of charged $B$
mesons, produced with high yield at LHC, to measure the absolute
branching fraction for the decay $\Lambda_c$ 
into $K p \pi$ - or any other observable decay mode - 
also in a model independent manner. The proposed
method
has the additional advantage of being applicable 
in a hadron collider environment since it does not require the
reconstruction of the complete event.
The paper is organised as follows. In Section 2 a description is 
given of the principle at the basis of
the proposed method and
in Section 3, by means of a simulation, 
the relations imposed
by kinematics are exploited. In Section 4 
selection efficiencies are evaluated, using 
the geometrical setup and quoted performance 
of the LHCb detector \cite{LHCb}, 
%an approximate experimental setup, 
to demonstrate the intrisic
feasibility of the proposed measurement. The effects
of non-resonant B-decays into the same final state
are also discussed. In Section 5 
the accuracy achievable with current data or with data available in the near
future at LHC is evaluated.

\section{Principle of $\Lambda_c$ reconstruction}

\begin{figure}[bt!]
\begin{center}
\includegraphics[width=8cm]{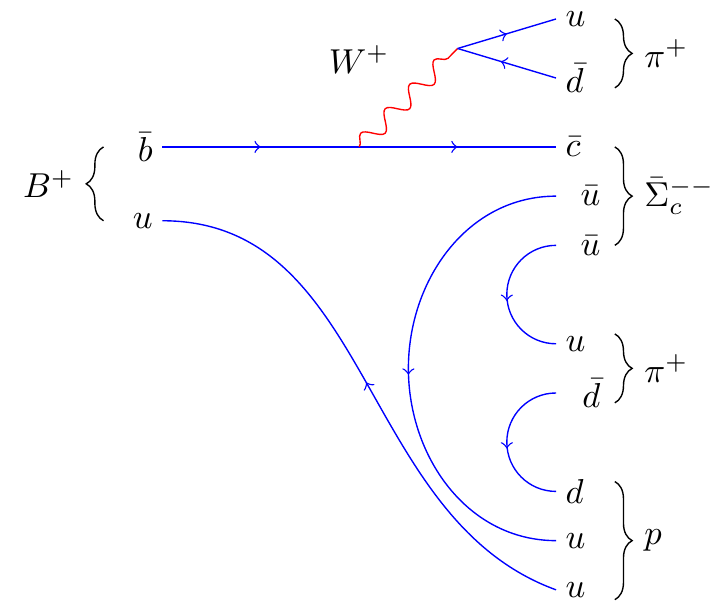}
\end{center}
\caption{Scheme of the decay \bdecay.}
\label{fig:decay}
\end{figure}

Even though the measured branching fraction of 
the decay \bdecay is only $2.8\cdot 10^{-4}$ \cite{PDG}
the abundant production of charged B-mesons at LHC makes it possible to
obtain samples containing O($10^7 - 10^8$) 
decays of this type. The method proposed in this
paper takes advantage of the decay chain \bdecaychain
whose diagram is shown in Fig. \ref{fig:decay},
\noindent
to measure absolute $\Lambda_c$ branching fractions
\begin {footnote}
{Unless otherwise indicated, charge conjugation is implicitely 
assumed throughout.}
\end {footnote}
.
The principle which the method relies upon is based on the kinematics
of the decay itself and is summarised in the following.
Assume that in the decay all charged particles are observed with the
exception of the $\Lambda_c$. If the direction of flight of
the $B$-meson is known, it is possible to infer the existence
of the $\Lambda_c$ and to determine its momentum 
\begin {footnote}
{Up to a quadratic ambiguity.}
\end {footnote}
 without observing its decay,
thus generating an unbiased sample of $\Lambda_c$'s 
in which one would search for the mode whose branching fraction is
sought to be measured.

The decay vertex of the $B^+$ ($B^-$) 
is identified by the presence of four charged particles 
(4-prong decay), 
namely $p \pi^+ \pi^+ \pi^-$ ($\bar{p} \pi^- \pi^- \pi^+$), 
having a total charge of $+2$($-2$). The $B$ meson direction of flight
is determined from the line joining the production (primary)
and decay vertices.

Furthermore, the pion whose sign of charge is opposite 
to that of the three remaining particles
certainly originates from the decay  of $\Sigma_c$ 
through strong interactions and,
in what follows, it will be referred to as pion from $\Sigma_c$,
$\pi_\Sigma$. In addition the presence of a proton would serve the 
purpose of tagging the decays of interest in experiments with good particle 
identification.

The B-meson decay vertex will be measured with an accuracy which 
depends on the experiment and is
separated from the production vertex by
a distance which would depend on the momentum spectrum of the B. 
The separation between the production and decay vertex and hence the
direction of flight of the B-meson will be measured with 
an experiment dependent accuracy as well. These factors will be taken
into account in Section 4.

Let $\hat{e}_B$ be a unit vector in the $B$ direction of flight
and $P_4 = (E_4, \textbf{p}_4)$
($P_4^{*}$) 
be the resultant four-momentum of all charged
particles at the B-decay vertex - not including the $\Lambda_c$ - in the
laboratory frame ($B$ rest frame).
$P_3$ and $P_3^{*}$ are 
the corresponding quantities 
of the three like-sign particles at the same vertex.
Let the invariant mass of the two systems be $M_4$
and $M_3$ respectively and
$\gamma$ be the Lorentz $\gamma$-factor of the decaying $B$-meson.

Assuming that a $\Lambda_c$ is the only missing particle in the decay,
through simple algebra, it can be shown that the following
two solutions are obtained for $\gamma$ 
depending on whether the system of four-particles 
moves forward or backward in the $B$ rest frame

\begin{equation}
\gamma_{1,2} = \frac{E_{4}\cdot E_{4}^* \mp |\vec{\textbf{p}}_{4}^{L}|\cdot|\vec{\textbf{p}}_{4}^{*L}|}{M_{4}^{2} + |\vec{\textbf{p}}_{4}^{*T}|^{2}} 
\label{eqn:gamma}
\end{equation}
\noindent
where $\vec{\textbf{p}}_4^T$ and $\vec{\textbf{p}}_4^L$ are the transverse 
and longitudinal momentum with respect to the B-flight direction
and $E_4^*$, the energy of the system of four particles
in the $B$ rest frame, is determined by the relation

$$E_{4}^* = \frac{M_{B}^{2} - M_{\Lambda_c}^{2} + M_{4}^{2}}{2M_{B}}$$

\noindent
Hence $E_B = \gamma \cdot M_B$ and 
$\vec{\textbf{P}}_B = \sqrt{(E_B^2 -M_B^2)} ~ \hat{e}_B$,
since the B flight direction in known.
The $\Lambda_c$ four-momentum will be determined
by imposing conservation of energy and momentum,
$P_\Lambda = P_B - P_4$, and therefore, if the $\Lambda_c$ 
truly originates from
a $\Sigma_c$ decay, its momentum would be such
that the combination
$(P_\Lambda + P_{\pi_\Sigma})^2$
must be equal to the $\Sigma_c$ mass squared.

This should result in a peaking of the mass
distribution around the true value of the $\Sigma_c$ mass
when the correct choice for
the Lorentz $\gamma$ factor has been made. 

In this manner, an unbiased sample of $\Lambda_c$ could be 
selected without actually observing the decay products of that particle
and therefore  
it would be sufficient to identify within it the presence of
decays into $p K \pi$ to measure the absolute branching fraction.

\section{Feasibility of the proposed method}

To demonstrate the viability of the proposed method, $pp$ interactions were
generated at centre of mass energy of $14~ TeV$, using the PYTHIA
generator \cite{Pythia}. $B^+$, produced over the whole solid
angle, were forced to decay in the channel of interest \bdecay
using the software package EVTGEN \cite{EvtGen}. Different
samples in which the decay was of the non-resonant types \nonresdecay
and \nonresdecpi
were also generated to investigate possible kinematical
variables which allow to separate the different decays.

There are indeed specific experimental advantages 
in using the suggested decay chain, some of which can be exploited
by detectors with
excellent particle identification, namely:

i) The four-prong decay vertex has charge $+2$. 
It is therefore relatively easy to identify and, in real experimental
conditions, would help in reducing background from decays
of particles other than the $B^+$.

ii) There is a proton at this vertex and therefore it can be efficiently 
identified to tag the $B^+$ decay.
Furthermore, the charge of this proton is opposite to that 
of the proton from $\Lambda_c$ decays and therefore no bias is
introduced from a specific $\Lambda_c$ decay mode.

iii) The pion from the $\Sigma_c$ has sign of charge opposite to that of
 the other three particles and therefore it can be unambiguously distinguished.

iv) Conditions that events lie within
 kinematical boundaries can be applied in the selection
to separate decays that occur through the resonance $\Sigma_c$ from the
non-resonant mode \nonresdecay, 
which has a branching fraction about 8 times larger, or from 
\nonresdecpi. In fact, if the final state $\bar{\Lambda}_c^- \pi^- p \pi^+ \pi^+$ 
is reached via the resonance $\bar{\Sigma}_c^{--}$, 
$M_4$ should have values between the minimum

$$(M_{4}^{2})_{min} = \huge{(}E_{3}^{'} + E_{\pi_{\Sigma}}^{'})^{2} - (\sqrt{E_{3}^{'2}-M_{3}^{2}}+\sqrt{E_{\pi_{\Sigma}}^{'2} - m_{\pi_{\Sigma}}^{2}}\huge{)}^{2} $$

\noindent
and the maximum

%\begin{equation}
$$(M_{4}^{2})_{max} = (E_{3}^{'} + E_{\pi_{\Sigma}}^{'})^{2} - (\sqrt{E_{3}^{'2}-M_{3}^{2}}-\sqrt{E_{\pi_{\Sigma}}^{'2} - m_{\pi_{\Sigma}}^{2}})^{2} $$
%\end{equation}   
\noindent
where $E_{3}^{'}$ and $E_{\pi_{\Sigma}}^{'}$ are the energies,
in the $\Sigma$ rest frame,
of the system of three-particles and of the $\pi_\Sigma$, respectively.

\begin{figure}[htb!]
%\begin{center}
\centering
\subfigure[Non resonant decay \nonresdecay]{%
\includegraphics[width=7cm]{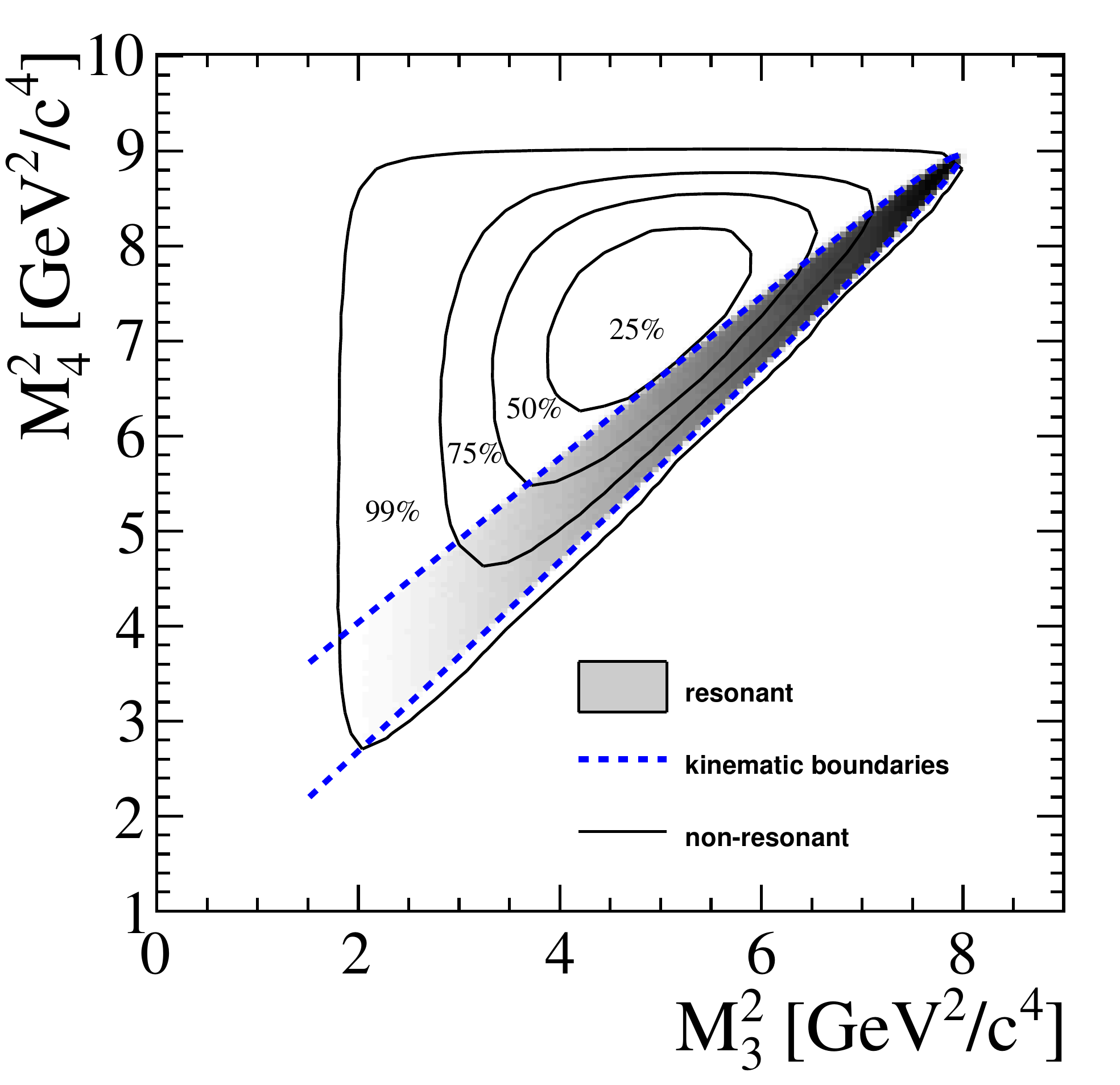}
 \label {fig:subfigure1}}
\subfigure[Non resonant decay \nonresdecpi]{%
\includegraphics[width=7cm]{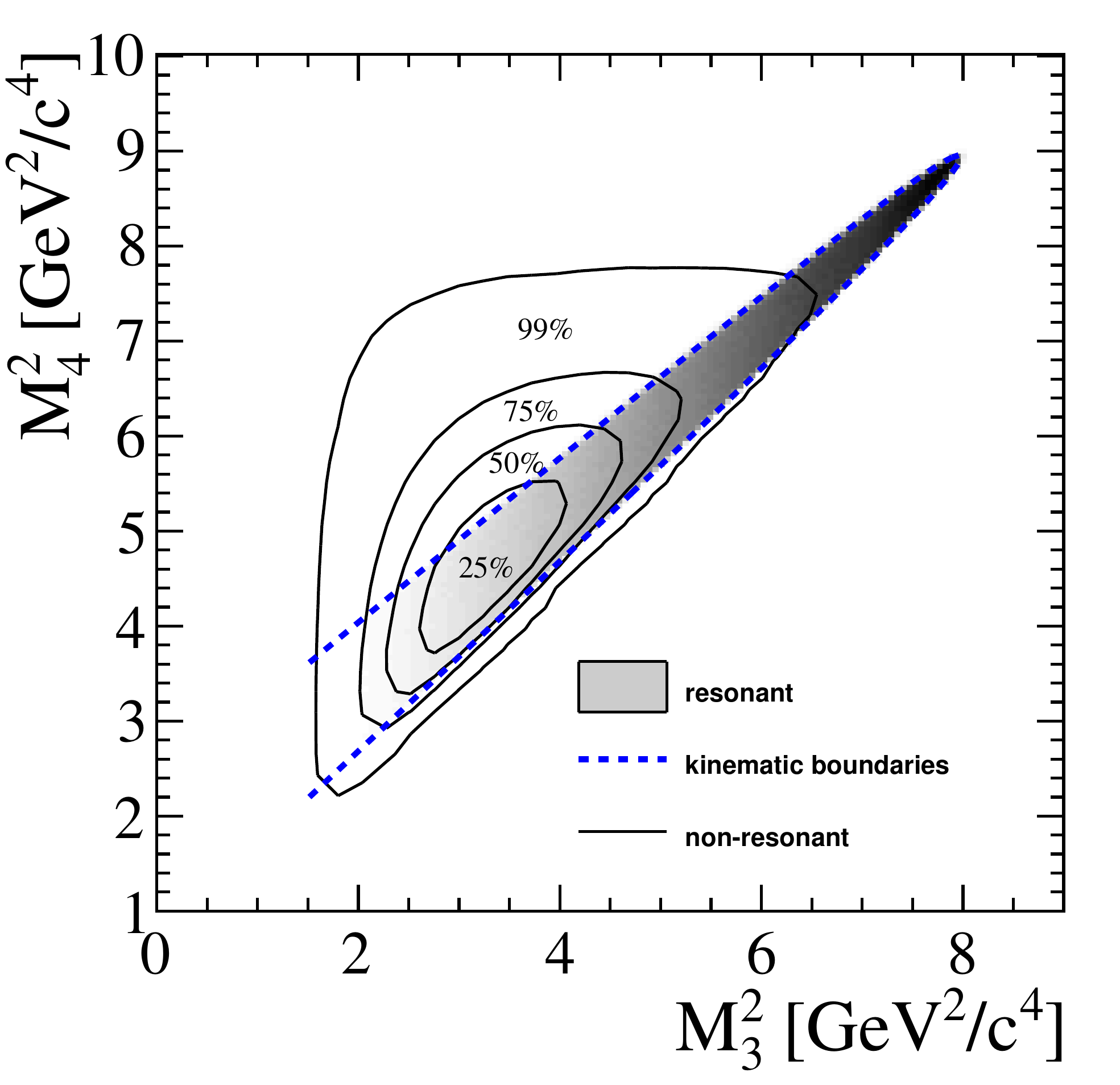}
 \label {fig:subfigure2}}
%\end{center}
%\vskip -0.5cm
\caption{Distribution of the 4-particle invariant mass squared versus 
3-particle invariant mass squared in resonant \bdecay and non-resonant \nonresdecay
and \nonresdecpi decays. The dashed line represents the kinematics 
boundaries for the resonant decay, while within the contour lines 
are contained the indicated fractions of non-resonant decays.}
\label{fig:M3M4}
\end{figure}

Fig.\ref{fig:M3M4} shows the kinematic boundaries 
defined above in the {\it Dalitz-plane}
($M^2_4 - M^2_3$).
Decays in which a $\Sigma_c$ is present fall
within these boundaries and are shown by the shaded area, while
within the contour lines are contained the indicated
fractions of non-resonant decays \nonresdecay (a) and \nonresdecpi (b).
The fraction of such decays within the kinematic 
boundaries is $(21.98 \pm 0.02)\%$ 
for \nonresdecay and $(50.46 \pm 0.03)\%$ for the \nonresdecpi mode.
In Section 4 a quantitative estimate of the contribution from these
decay modes will be given, once further selections have been applied. 
At this stage it is
sufficient to observe that requiring a minimum value of $M_3^2$ and $M_4^2$
would be effective in reducing the fraction of non resonant decays,
in particular of \nonresdecpi whose branching fraction is not measured
at present and only an upper limit exists \cite{PDG}.

\begin{figure}[htb!]
\begin{center}
\includegraphics[width=12cm]{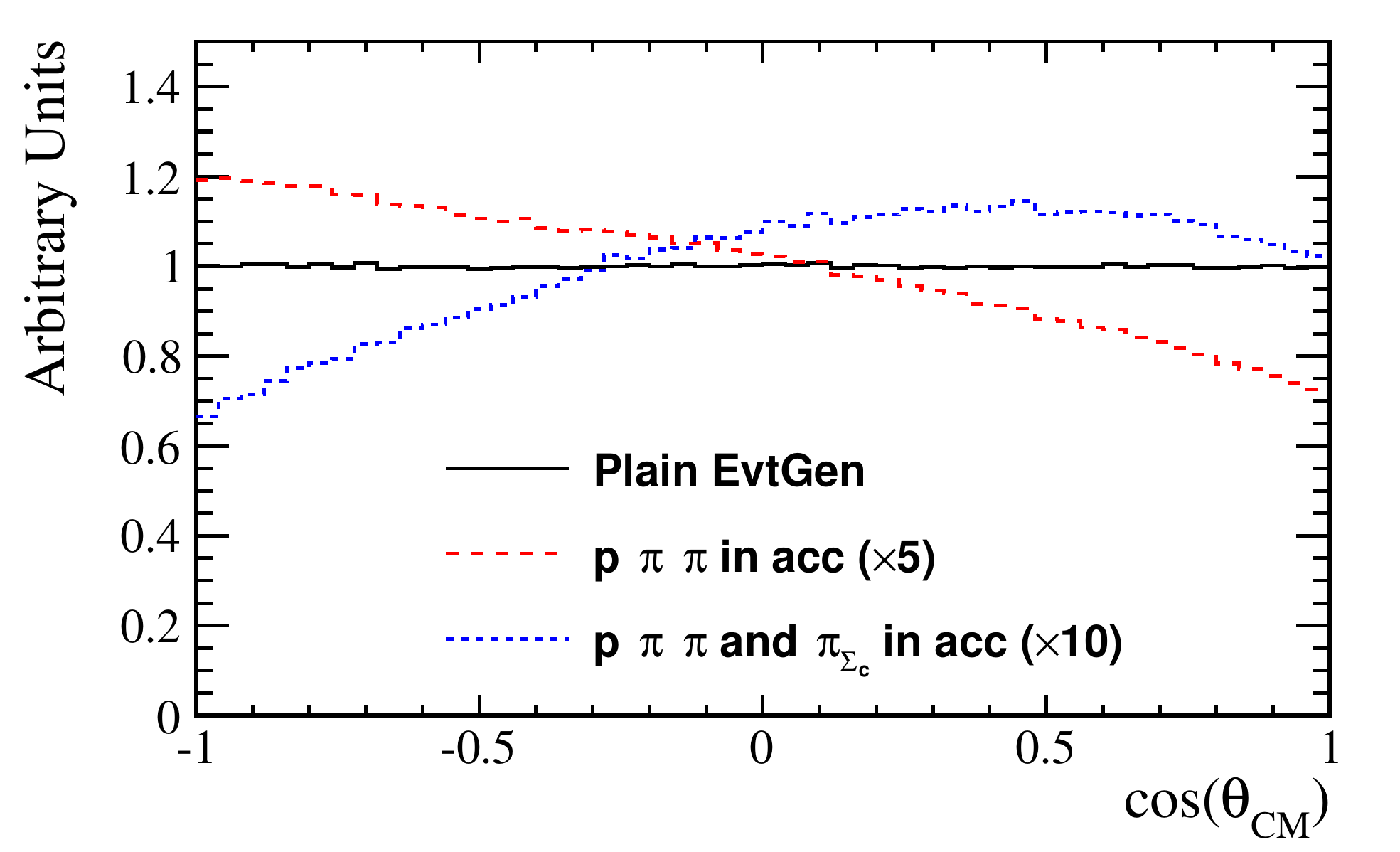}
\end{center}\label{fig:costheta}
\vskip -0.5cm
\caption{Angular distributions of the
$\Lambda_c$ baryon in the $B$ rest frame before and after applying
selection criteria}
\end{figure}

Detector acceptance was simulated in a simple manner, by assuming
that particles with momentum greater than
$2~GeV/c$ and within the pseudo-rapidity range $2 < \eta < 4.5$
would be detectable, as it would be approximately the case 
in the LHCb experiment, whose simplified setup will be used to
estimate efficiencies.
In the decay \bdecaychain, the spatial distributions of particles
are generated according to phase space, therefore, at generator level,
there is no preferential direction for the $\Lambda_c$ in the $B$ rest 
frame. 
However, requiring that the decay products - other than the $\Lambda_c$ -
be within a detector acceptance, introduces asymmetries of order of $20\%$.
This is shown in Fig. \ref{fig:costheta}, where the distributions of the
cosine of the angle between the direction of the momentum of the
$\Lambda_c$ in the $B$ rest frame and the $B$ direction of 
flight in various situations
are compared.

There are indeed also losses
introduced by this selection and these will be included in
the efficiencies discussed in Section 4. 
At this stage it is sufficient to observe that
requiring that the three like-sign particles be in acceptance favours
slightly backward-going $\Lambda_c$'s, while the further requirement
that the pion from $\Sigma_c$ be measurable, i.e. within
the geometrical acceptance as well,  would preferentially
select forward-going $\Lambda_c$, as shown by the dashed lines in 
Fig. \ref{fig:costheta}. 

This selection of different regions of phase space will have
an effect on the preferred value for the Lorentz $\gamma$ factor.
In fact, with the definition given in Eq. \ref{eqn:gamma}, 
$\gamma_1$ is always smaller than $\gamma_2$ and therefore
chosing the first over the latter selects $B$'s of lower energy.
If for instance experimental acceptance is such that forward going $\Lambda_c$ 
are slightly favoured, then $\gamma_2$ would be
more often the correct solution.
On the other hand, additional selections - that might prove necessary
when dealing with the real experimental conditions - are likely
to change the favoured value.
However this would have no effect on the proposed measurement,
as long as no bias is introduced by the decay mode
of the $\Lambda_c$, which is the case here since the $\Lambda_c$ decay
products are not and will not be considered in the selection.

\begin{figure}[htb!]
\begin{center}
\includegraphics[width=12cm]{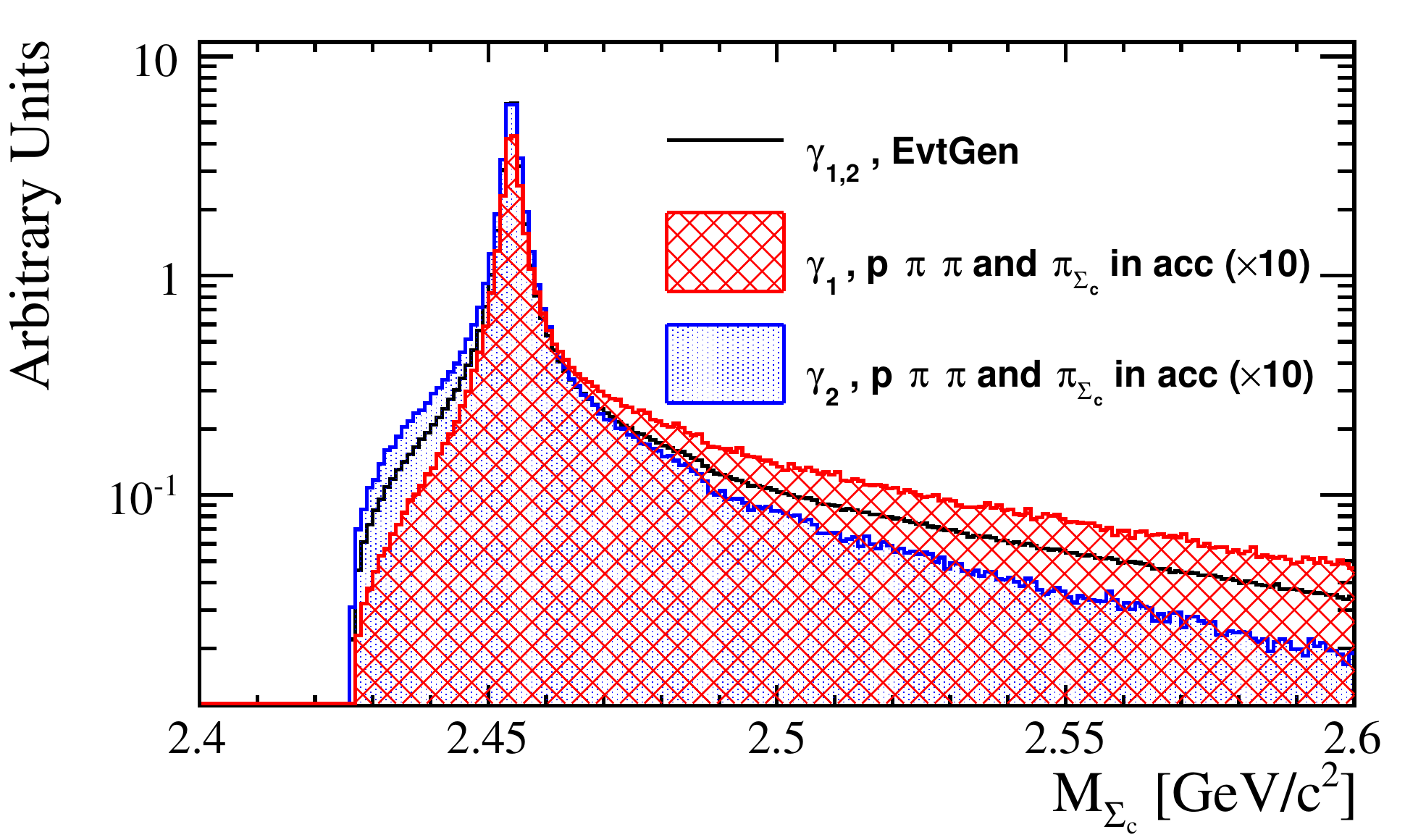}
\end{center}
\vskip -0.5cm
\caption{Distributions of the reconstructed $\bar{\Sigma}_c^{--}$ mass before
and after applying selection criteria.}
\label{fig:sigmagen}
\end{figure}

For each event, the two values of the Lorentz $\gamma$ factor
given by Eq. \ref{eqn:gamma}
are computed using true quantities at generator level. 
$\sqrt{(P_\Lambda + P_{\pi_\Sigma})^2}$ is then computed assuming, as solution,
either $\gamma_1$ or $\gamma_2$.
The mass distribution obtained in this manner
is shown in Fig. \ref{fig:sigmagen}
as solid histogram. As already mentioned there is complete
symmetry forward-backward at this stage and therefore chosing either
solution, $\gamma_1$ or $\gamma_2$, leads to the same
result. As expected, a peak around the $\Sigma_c$ mass is observed.
Its width is affected by the cases in which the wrong
solution for $\gamma$ was chosen, yet the result was close to its true value
and, as a consequence, a value 
about $10\%$ larger than expected is observed
\begin {footnote}
{At this level, if the correct solution were chosen,
one would expect to obtain a width of $2.6~MeV/c^2$
for the $\Sigma_c$, which is the value coded in the Montecarlo.}
\end {footnote}
.
\noindent
  
Using either value of $\gamma$ and requiring  
$2.44~GeV/c^2 < M_\Sigma < 2.47~GeV/c^2$, 
would allow a selection of a sample
of $\Lambda_c$ whose size is about 64\% of the original sample of
generated events.
On the other hand chosing particles that fall within the geometrical
acceptance would produce different mass distributions for the two 
choices of $\gamma$, as illustrated in Fig. \ref{fig:sigmagen}
with slightly different efficiencies which will be discussed 
quantitatively in Section 4. 

Even though the choice of $\gamma_1$ or of $\gamma_2$ is
irrelevant since either one would allow the selection of
an unbiased sample of $\Lambda_c$'s, efficiencies are 
different and this would affect the size of the final sample
and ultimately the statistical accuracy.
To minimize the statistical error on the measurement, it would
be required in addition that the decay products of the $\Lambda_c$ be 
detected and hence it is the number of observable decays
that ought to be maximized.

\section{Effects of experimental resolution and efficiencies}

\begin{figure}[htb!]
\begin{center}
\includegraphics[width=12cm]{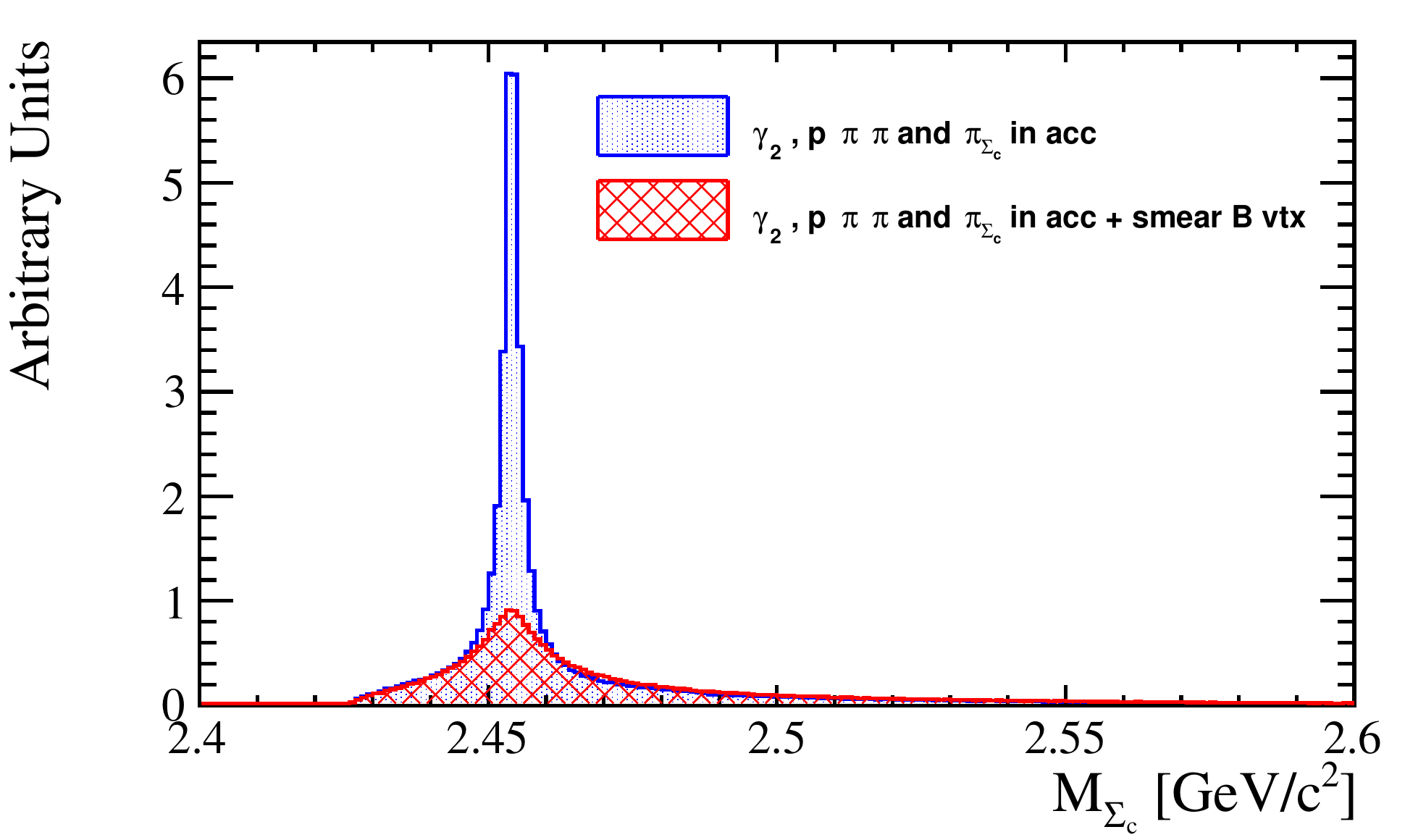}
\end{center}
\label{fig:sigma_smear}
\vskip -0.5cm
\caption{Distributions of the reconstructed $\bar{\Sigma}_c^{--}$ mass
before and after applying resolution smearing.}
\end{figure}

Experimental resolution on the determination of
momenta of particles and of positions of vertices so far have not 
been taken into account. To estimate its effect, a simplified
geometrical setup of the LHCb experiment was used in the Montecarlo
simulation. 
Therefore the momentum-dependent smearing 
quoted in ref. \cite{LHCbP} (which corresponds to $\delta P/P \sim 0.4\%$ for
a particle with $P = 10~GeV/c$) was applied to particle momenta and
gaussian smearings of $\sigma_z = 400 \mu m$ along the beam direction and 
of $\sigma_{T} = 35 \mu m$ in the transverse beam direction were applied
to the position of the $B$ decay vertex \cite{LHCbVX}.
It was found that, of the two,
the error on the vertex position has the largest effect
since it enters in the determination of the $B$ direction of flight.
The effect is shown in Fig. \ref{fig:sigma_smear}
where the $M_{\Sigma_c}$ distribution is displayed for the
$\gamma_2$ solution and all four-particles within the geometrical acceptance,
assuming the true $B$-direction or that obtained having
applied only the smearing on the $B$-decay vertex.
The result is essentially unchanged when the momentum smearing is also applied.
The effect of the experimental resolution is to reduce
the event sample to about $45\%$ of its original size and
it is mostly due to the fact that the transverse momentum
relative to the {\it measured} B-direction exceeds the
maximum value allowed by kinematics and hence no acceptable solutions
for $\gamma$ are found.

 Tables \ref{table:efficiency1} and \ref{table:efficiency2} 
summarise the effects
of the selections listed below and applied to the \bdecay sample as well
as to the {\it {non-resonant}} samples,
i.e. to the samples of events in which the 
final state $\Lambda_c \pi p \pi \pi$,
with or
without the presence of an additional neutral pion,
is reached
directly from B decay and not via the resonance $\Sigma_c$
for the two possible choices of $\gamma$:

\begin{itemize}
\item {Selection {\bf A}:} $(M_4^2)_{min} < M_4^2 < (M_4^2)_{max}$
\item {Selection {\bf B}:} Mass of the $\Sigma_c$, computed as
$\sqrt{(P_\Lambda + P_{\pi_\Sigma})^2}$, is within $\pm 15~MeV/c^2$ from
its nominal value of $2.455~GeV/c^2$.
\item {Selection {\bf C}:} $p \pi \pi$ like-sign within geometrical acceptance.
\item {Selection {\bf D}:} $\pi_\Sigma$ within geometrical acceptance.
\item {Selection {\bf E}:} Smearing is applied to B-decay vertex.
\item {Selection {\bf F}:} $M_3^2 > 4 (GeV/c^2)^2$ and $M_4^2 > 6 (GeV/c^2)^2$
\end {itemize} 

The criteria were applied in order, i.e. each selection implies
that all the preceeding conditions were satisfied.

\begin{table}[hbpt]
\begin {center}
  \caption{Accepted solution $\gamma_1$ - Fraction of events kept}
\begin{tabular}{|l|l|l|l|}
\hline
Condition&\bdecay &\nonresdecay & \nonresdecpi \\
\hline 
Selection {\bf A} & $1$ & $(21.98\pm 0.02)\cdot 10^{-2}$ & $(50.46\pm0.03) \cdot 10^{-2}$  \\ 
      \hline
Selection {\bf B} & $(62.19\pm 0.03)\cdot 10^{-2}$ & $(1.986\pm 0.005)\cdot 10^{-2}$ & $(2.493\pm0.006) \cdot 10^{-2}$  \\ 
      \hline
Selection {\bf C} & $(6.501\pm 0.009)\cdot 10^{-2}$ & $(2.00\pm 0.02)\cdot 10^{-3}$ & $(2.39\pm0.02) \cdot 10^{-3}$  \\ 
      \hline
Selection {\bf D} & $(2.554\pm 0.006)\cdot 10^{-2}$ & $(6.89\pm 0.09)\cdot 10^{-4}$ & $(3.03\pm0.06) \cdot 10^{-4}$  \\ 
      \hline
Selection {\bf E} & $(1.055\pm 0.004)\cdot 10^{-2}$ & $(3.79\pm 0.07)\cdot 10^{-4}$ & $(2.76\pm0.06) \cdot 10^{-4}$  \\ 
      \hline
Selection {\bf F} & $(8.52 \pm 0.03)\cdot 10^{-3}$ & $(2.96 \pm 0.06 )\cdot  10^{-4}$ & $ (1.07 \pm 0.04)\cdot 10^{-4}$  \\ 
      \hline
    \end{tabular}
    \label{table:efficiency1}
  \end{center}
\end{table}

\begin{table}[hbpt]
  \begin{center}
  \caption{Accepted solution $\gamma_2$ - Fraction of events kept}

\begin{tabular}{|l|l|l|l|}
\hline
Condition&\bdecay &\nonresdecay & \nonresdecpi \\
\hline 
Selection {\bf A} & $1$ & $(21.98\pm 0.02)\cdot 10^{-2}$ & $(50.46\pm0.03) \cdot 10^{-2} $  \\ 
      \hline
Selection {\bf B} & $(64.02\pm 0.04)\cdot 10^{-2}$ & $(2.014\pm 0.005)\cdot 10^{-2}$ & $(2.581\pm0.006) \cdot 10^{-2}$  \\ 
      \hline
Selection {\bf C} & $(5.784\pm 0.009)\cdot 10^{-2}$ & $(1.94\pm 0.02)\cdot 10^{-3}$ & $(2.17\pm0.02) \cdot 10^{-3}$  \\ 
      \hline
Selection {\bf D} & $(3.403\pm 0.007)\cdot 10^{-2}$ & $(1.29\pm 0.01)\cdot 10^{-3}$ & $(1.79\pm0.02) \cdot 10^{-3}$  \\ 
      \hline
Selection {\bf E} & $(1.536\pm 0.004)\cdot 10^{-2}$ & $(8.3\pm 0.1)\cdot 10^{-4}$ & $(1.43\pm0.01) \cdot 10^{-3}$  \\ 
      \hline
Selection {\bf F} & $(1.117 \pm 0.004)\cdot 10^{-2}$ & $(5.62 \pm 0.08 )\cdot  10^{-4}$ & $ (3.49 \pm 0.07)\cdot 10^{-4}$  \\ 
      \hline
    \end{tabular}
    \label{table:efficiency2}
  \end{center}
\end{table}

From the tables it can be concluded that
efficiencies at the percent level for the
indicated selections are obtained and that, as expected,
$\gamma_2$ would be the favoured solution when requiring that  
$\pi_\Sigma$ be within geometrical acceptance
(i.e. having pseudo-rapidity in the range $2 < \eta < 4.5$, as
previously defined).

\begin{figure}[htb!]
\begin{center}
\includegraphics[width=12cm]{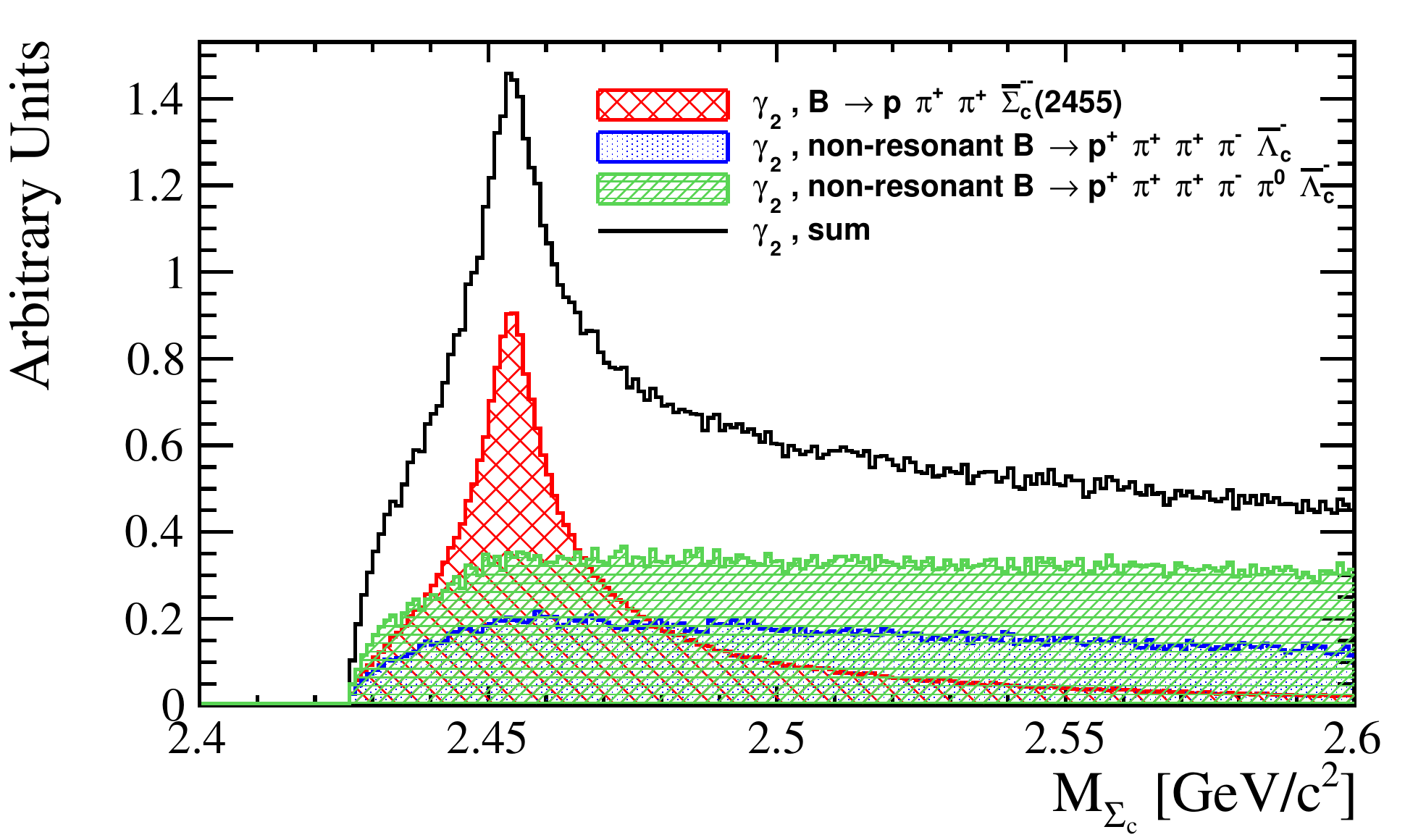}
\end{center}
\label{fig:sigma_smear_nonres}
\vskip -0.5cm
\caption{Distributions of the reconstructed $\bar{\Sigma}_c^{--}$ mass
in resonant (\bdecay) and non-resonant decays (\nonresdecay and \nonresdecpi).}
\end{figure}

As it was also expected,
the requirements imposed on kinematics and detector geometry
are less effective for the non-resonant channels, the
efficiencies being smaller by a factor of
about 20. This reduction will be partly compensated for by the larger
B-decay branching fraction into these channels. Quantitatively
this is shown in Fig. \ref{fig:sigma_smear_nonres} where the $\Sigma_c$ mass
distribution is displayed for the resonant and non-resonant channels
properly weighted with the B-decay branching fractions 
and the efficiencies
quoted in Table \ref {table:efficiency2} for the selection {\bf A}. 
For the decay \nonresdecpi, since only an upper
limit exists for the branching fraction, it was assumed to be
equal to that of \nonresdecay.
In the mass range $2.44-2.47 ~GeV/c^2$,considered as the signal region,
the total non-resonant fraction, with the above assumption, is
$\sim 36\%$ of the total.
In real experimental conditions this signal would be superimposed 
to a combinatorial background, therefore
it would not be meaningful at this stage to extract a function describing its shape.

As already pointed out, it is the number of observable decays
that should be maximised. Table \ref{table:kppi_eff} shows,
for the decay $\ldecay$, the fraction of events,
satisfying the indicated selections,
in which the decay products of the $\Lambda_c$ are also within geometrical
acceptance, both for the resonant and non-resonant components.
 Efficiencies are of order of $\sim 40\%$ irrespective of
the choice of $\gamma$ and depend weakly on whether the $\Lambda_c$
is produced directly from B decay or through the $\Sigma_c$ resonance.
This was to be expected, since essentially only geometrical factors
enter in the determination of these fractions.

\begin{table}[hbpt]
\begin{center}
  \caption{Fraction of events for the indicated selections in which the
 $\Lambda_c$ decay products are also within geometrical acceptance}

\begin{tabular}{|l|l|l||l|}
\toprule 
Condition &\bdecay & \nonresdecay & \nonresdecpi \\
\hline 
Selection {\bf D} & & & \\
and Solution $\gamma_1$
 &$39.82 \pm 0.12$ & $ 42.6 \pm 0.8 $ & $40.9 \pm 1.1$  \\ \hline
Selection {\bf E} & & & \\
and Solution $\gamma_1$
& $39.17 \pm 0.18$ & $40.7 \pm 1.0$ & $41.5 \pm 1.1$ \\
\hline 
Selection {\bf F} & & & \\
and Solution $\gamma_1$
 &$39.16 \pm 0.2$ & $ 41.4 \pm 1.1 $ & $42.3 \pm 1.9$  \\ \hline
\hline
Selection {\bf D} & & & \\
and Solution $\gamma_2$
 &$40.54 \pm 0.10$ & $ 41.4 \pm 0.5 $ & $39.5 \pm 0.4$  \\ \hline
Selection {\bf E} & & & \\
and Solution $\gamma_2$
& $40.03 \pm 0.15$ & $40.3 \pm 0.6$ & $39.8 \pm 0.5$ \\ 
\hline
Selection {\bf F} & & & \\
and Solution $\gamma_2$
& $40.04 \pm 0.18$ & $41.9 \pm 0.8$ & $41.8 \pm 1.0$ \\
\hline 
    \end{tabular}
    \label{table:kppi_eff}
  \end{center}
\end{table}

\section{Results and Conclusions}

LHCb has measured a cross section of $38.9~\mu b$ for the production 
of charged B mesons within the experimental acceptance and with transverse
momentum in the range $0 - 40~GeV/c$ \cite{LHCb_xsec}, in $pp$ collisions
at centre of mass energy of $7~TeV$. This would correspond
to a cross section of $~193 \mu b$ over the whole solid
angle, assuming that B-mesons are produced as in PYTHIA generator.
Scaling the production cross-section with $\sqrt s$, this would
yield about $3.9 \cdot 10^{11}$ charged B-decays per $fb^{-1}$ of integrated
luminosity.

 Selecting, for the sake of illustration, the efficiencies corresponding
to the $\gamma_2$ solution in Table \ref{table:efficiency2}
and assuming that these do not have a strong
dependence on the pp centre-of-mass energy, using an integrated luminosity of
$3~fb^{-1}$ (currently available in the LHCb experiment at centre of
mass energies of $7$ and $8~TeV$),
with the measured branching fraction 
of $2.8 \cdot 10^{-4}$ for $\bdecay$, about
$2.5$ million decays of interest would be {\it reconstructed} 
within the detector:
a sizeable, unbiased sample of $\Lambda_c$ decays.

However, to obtain a more realistic estimate of the above number,
the effects of other selections which
is necessary to apply and have not been considered here,
should be included. 
These in general are functions of transverse momentum and pseudo-rapidity 
and efficiencies have typically the values quoted in Ref. \cite{LHCb_xsec1}.
Assigning the realistic value of $2\%$ to account for all the
effects not included in the simplified simulation,
using the efficiencies of Table \ref{table:kppi_eff} for the
$\Lambda_c$ decay products, about 1000 decays of the type
$\ldecay$ from the resonant sample
would be observed in the detector. The statistical
error therefore would 
be comparable with that of Belle \cite{Belle}, which
is the most precise currently available measurement.

The presence of large backgrounds could spoil the effectiveness of the
proposed method.
These can be estimated only through a detailed and
complete simulation using a specific detector and it is beyond the scope
of this article.
However it should be observed that the choice 
was made - over other topologies which would have enlarged the sample -
of a $B$-decay vertex of
charge $\pm 2$, with a well defined topology,
separated from the primary $pp$ interaction vertex, and containing
a well identified proton,
with the intent of reducing combinatorial background.
Backgrounds from specific decays of B-mesons have been
considered. In particular, as seen in Tables \ref{table:efficiency1} and
\ref{table:efficiency2} and in Fig. \ref{fig:sigma_smear_nonres},
non-resonant decays \nonresdecay and \nonresdecpi, contribute
less than $40\%$ to the total number.
The decay \nonresdecpi with a missing neutral pion may be considered
representative of the class of decays  with topology 
identical to that of interest when one or more particles are missing. 
The validity of this statement was verified by considering
in the simulation also the decay \bodecay, which would mimic the signal
when missing a $\pi^-$.
A fraction of about $3.7~10^{-4}$ of events
of this type were found for selection {\bf F},
in the mass region of interest.
As expected, the retained fraction is
similar to that accepted for the decay \nonresdecay.
The cross section for $B^\circ$ production is similar
to that for charged B's \cite{LHCb_xsec}.
However for the branching fraction of the decay considered here only an 
upper limit exists and therefore it would contribute to background
at most $\sim 30\%$ of the resonant signal.

Systematic effects are experiment dependent as well and therefore 
can not be properly estimated in this paper.
However it is worth noticing that the method relies on {\it counting}
the observed number of \ldecay decays in a sample whose selection
does not rely upon observation of the $\Lambda_c$ decay products.
Therefore most of the systematics would cancel when taking the ratio.
The line shape used in the fit to determine the size of the
initial, unbiased sample would be taken from Montecarlo simulation. 
The effects of the uncertainty on this shape would not cancel 
however and would most likely be the main source of systematic error.
Particle identification (proton in particular) would be used
to identify the $\Lambda_c$ decay mode of interest. This
would affect only
the detection efficiency listed
in Table \ref{table:kppi_eff} and therefore it would not cancel in the
ratio and would become important if statistics were limited.

The overall result could be improved by devising  dedicated, more
efficient selections at trigger level and when more data and at higher
centre of mass energy become available. 
Furthermore the decays from
$\Lambda_c$'s originating from the non-resonant channels could be
added to the sample, since their detection efficiency is similar to
that for $\Lambda_c$ coming from 
the resonant channel and therefore large corrections would not
be required.

%\Section*{Acknowledgements}

%\clearpage

\end{document}